\documentclass[aip,apl,amsmath, amssymb, reprint]{revtex4-1}

\usepackage{graphicx}
\usepackage{dcolumn}
\usepackage{bm}
\usepackage{color}

\newcommand{\lessgtrsim}[1]{\raisebox{0.05em}{\parbox[][][t]{0.7em}{\scalebox{0.8}[0.7]{$#1$}}}}

\newcommand{\Ahfs}[0]{\mathcal{A}_{\textrm{hfs}}}
\newcommand{\muB}[0]{\mu_{\textrm{B}}}
\mathchardef\bmu= "7916

\begin{document}


\title{A Remotely Interrogated All-Optical $^{87}$Rb Magnetometer}

 \author{B.\ Patton}
 \email{bpatton@berkeley.edu}
    \address{Department of Physics, University of California, Berkeley, CA 94720-7300}

 \author{O.\ O.\ Versolato}
    \address{University of Groningen, Kernfysisch Versneller Instituut, NL-9747 AA Groningen, The Netherlands}
    \altaffiliation{Current address: Max-Planck-Institut f\"{u}r Kernphysik, Heidelberg, Germany}

 \author{D.\ C.\ Hovde}
    \address{Southwest Sciences, Inc., Cincinnati, OH 45244}

 \author{E.\ Corsini}
    \address{Department of Physics, University of California, Berkeley, CA 94720-7300}

 \author{J.\ M.\ Higbie}
    \address{Department of Physics and Astronomy, Bucknell University, Lewisburg, PA 17837}
%

 \author{D.\ Budker}
    \address{Department of Physics, University of California, Berkeley, CA 94720-7300}
    \address{Nuclear Science Division, Lawrence Berkeley National Laboratory, Berkeley, CA 94720}
\date{\today}

\begin{abstract}

Atomic magnetometry was performed at Earth's magnetic field over a free-space distance of ten
meters. Two laser beams aimed at a distant alkali-vapor cell excited and detected the $^{87}$Rb
magnetic resonance, allowing the magnetic field within the cell to be interrogated remotely.
Operated as a driven oscillator, the magnetometer measured the geomagnetic field with
\lessgtrsim{\lesssim}3.5\,pT precision in a $\sim$2\,s data acquisition; this precision was
likely limited by ambient field fluctuations. The sensor was also operated in self-oscillating
mode with a 5.3\,pT/$\sqrt{\textrm{Hz}}$ noise floor. Further optimization will yield a
high-bandwidth, fully remote magnetometer with sub-pT sensitivity.

\end{abstract}

\maketitle


Shortly after the inception of atomic magnetometry, alkali-vapor magnetometers were being used to
measure the Earth's magnetic field to unprecedented precision. During the same era, Bell and
Bloom first demonstrated all-optical atomic magnetometry through synchronous optical
pumping~\cite{Bell_PRL61}.  In this approach, optical-pumping light is frequency- or
amplitude-modulated at harmonics of the Larmor frequency $\omega_{\textrm{L}}$ to generate a
precessing spin polarization within an alkali vapor at finite magnetic field. Although this
technique received considerable attention from the atomic physics community for its applicability
to optical pumping experiments, Earth's-field alkali-vapor atomic magnetometers continued to rely
on radiofrequency (RF) field excitation for several decades. Upon the advent of diode lasers at
suitable wavelengths, synchronously pumped magnetometers experienced a revival beginning the late
1980s. In recent years, advances in all-optical magnetometers using
amplitude-modulated\cite{Gawlik_APL06} and frequency-modulated\cite{Acosta_PRA06} light have
resulted in applications such as nuclear magnetic resonance detection\cite{Bevilacqua_JMR09},
quantum control experiments\cite{Pustelny_PRA11}, and chip-scale devices intended for spacecraft
use\cite{Korth_JHAPL10}.


All-optical magnetometers possess several advantages over devices which employ RF coils.
RF-driven magnetometers can suffer from cross-talk if two sensors are placed in close proximity,
since the AC magnetic field driving resonance in one vapor cell can adversely affect the other.
All-optical magnetometers are free from such interference.  When operated in self-oscillating
mode~\cite{Bloom_AO62}, RF-driven magnetometers require an added $\pm90^{\circ}$ electronic phase
shift in the feedback loop to counter the intrinsic phase shift between the RF field and the
probe-beam modulation. In an all-optical magnetometer this same phase shift can be achieved
simply by varying the relative orientations of the pump and probe beam
polarizations~\cite{Higbie_RSI06}. Most importantly, all-optical magnetometers require no
physical connection between the driving electronics and the alkali-vapor cell. This allows
completely remote interrogation of the magnetic resonance in a faraway atomic sample.  Here we
describe a demonstration of remotely interrogated all-optical magnetometry

\begin{figure*}[t] \centering
\includegraphics[scale=1]{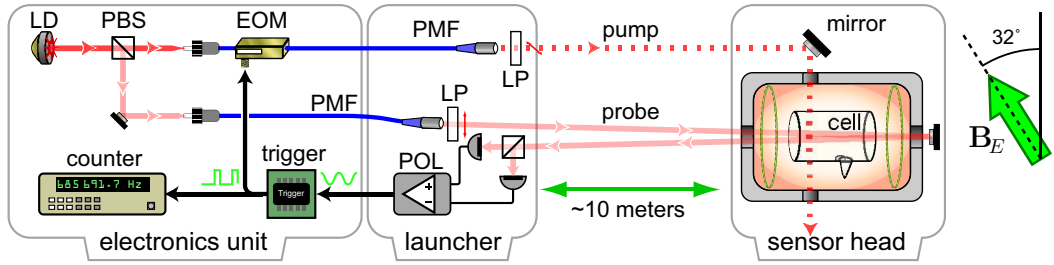}
{\caption[Remote magnetometry setup.]{\label{RemoteSetup} (Color online) Schematic of the
remote-detection magnetometer. The laser diode (LD) beam was split with a polarizing beamsplitter
(PBS) and coupled into two polarization-maintaining fibers (PMF). The pump beam amplitude was
modulated by a fiberized EOM.  At the launcher both beams were collimated, sent through linear
polarizers (LP), and aimed at the sensor head. The pump beam was polarized horizontally, creating
atomic alignment perpendicular to the ambient field.  The probe beam reflected off a mirror aimed
at a balanced polarimeter (POL) within the launcher. In self-oscillating mode (depicted), the
polarimeter output was conditioned to drive the EOM directly. In driven-oscillation mode, the EOM
was driven by a swept frequency source and the polarimeter output demodulated with a lock-in
amplifier. The Earth's field $\textbf{B}_E$ was independently measured to be 32$^{\circ}$ from
the vertical in the direction depicted.}}
\end{figure*}

A schematic of the remote-detection magnetometer is shown in Fig.~\ref{RemoteSetup}. The
unshielded sensor was similar to that described in Ref.~\cite{Higbie_RSI06} in that the pump and
probe beams were derived from a single laser whose frequency was stabilized by a dichroic atomic
vapor laser lock (DAVLL)~\cite{Yashchuk_RSI00}. The atomic sample consisted of an
antirelaxation-coated\cite{Balabas_PRL10} alkali-vapor cell containing enriched $^{87}$Rb and no
buffer gas; the longitudinal spin relaxation time of atoms within the cell was 1.2 s.  The laser
beams were carried from an optics and electronics rack to a launcher assembly via
polarization-maintaining optical fibers; the pump beam amplitude was modulated with a fiberized
Mach-Zender electro-optic modulator (EOM).  At the launcher, the collimated output beams were
linearly polarized and aimed at a sensor head placed ten meters away.  This assembly contained
the $^{87}$Rb cell within an enclosure heated to 34.5~$^{\circ}$C by a 1.7 kHz alternating
current flowing through counter-wrapped heating wires.  These wires comprised the only physical
contact between the experimental apparatus and the atomic sample.  In principle such heating is
not necessary, but it was employed here to boost the optical-rotation signal above electronic
interference from AM radio stations.

The probe beam traveled horizontally through the optical cell in a double-pass configuration,
reflecting off a mirror behind the cell and propagating back toward the launcher. There a
balanced polarimeter split the probe beam into orthogonal polarizations which were projected onto
two photodiodes, allowing optical rotation to be measured.  Synchronous optical pumping at
2$\omega_{\textrm{L}}$ created atomic alignment\cite{Yashchuk_PRL03,Note1} within the $^{87}$Rb
vapor which produced time-varying optical rotation of the probe polarization at
$\omega_{\textrm{L}}$ and 2$\omega_{\textrm{L}}$. The $\omega_{\textrm{L}}$ harmonic arises when
the field is tilted away from the direction of the probe beam propagation
vector\cite{Pustelny_PRA06}.  In the current experiment the ambient geomagnetic field pointed
32$^{\circ}$ from the vertical, leading to an optical rotation component at $\omega_{\textrm{L}}$
which was several times larger than the 2$\omega_{\textrm{L}}$ harmonic. Note that this field
orientation is far from optimal, since both harmonics exhibit zero amplitude when the magnetic
field is perpendicular to the probe beam propagation\cite{Pustelny_PRA06} (a configuration known
as a ``dead zone'').

The Zeeman shifts of the alkali ground-state sublevels (total electron spin $J = 1/2$) at a
magnetic field $B$ can be calculated from the Breit-Rabi equation\cite{OPA_Ed1}.
\begin{eqnarray}
E(F, m_F) & = & -\frac{\Ahfs}{4} - g_I \muB m_F B \nonumber \\ & \pm & \frac{\Ahfs
(I+\tfrac{1}{2})}{2}\sqrt{1+\frac{4 m_F x}{2I+1}+x^2},
 \label{BR01}
\end{eqnarray}
where $E$ is the energy of the ground state sublevel with quantum numbers $F$ and $m_F$ ($F = I
\pm \frac{1}{2}$ being the total angular momentum of the ground state and $I$ the nuclear spin),
$\Ahfs$ is the hyperfine structure constant, and the perturbation parameter $x$ is given by:
\begin{equation}
x \equiv \frac{(g_J + g_I)\muB B}{\Ahfs (I+\tfrac{1}{2})}. \label{xequals}
\end{equation}
Here $g_J$ and $g_I$ are the electron and nuclear $g$ factors, respectively\cite{Note2}. At low
magnetic fields, a linear approximation to Eq.~(\ref{BR01}) predicts a single resonance at
$\omega_{\textrm{L}}$ for all transitions with $\Delta m_F = 1$ and another resonance at
2$\omega_{\textrm{L}}$ for all $\Delta m_F = 2$ transitions. At Earth's field ($B_E$), these
resonances split into sets of resolved transitions due to higher-order corrections.  In the
present study the laser was tuned to address the $F=2$ ground-state hyperfine manifold of
$^{87}$Rb, yielding four resonances with $\Delta m_F = 1$ and three with $\Delta m_F = 2$. The
magnetometer is nominally designed to probe $\Delta m_F = 2$ resonances in order to reduce
systematic errors\cite{Patton_HE}, but in this study it could also be operated near
$\omega_{\textrm{L}}$ due to the field configuration.

\begin{figure}[b]
\includegraphics[scale=1]{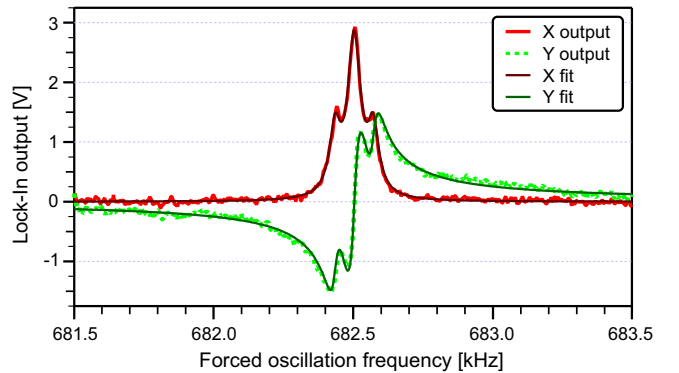}
{\caption[Lock-in data.]{\label{FO} (Color online) Driven-oscillation data recorded with the
remote magnetometer.  A spurious background has been subtracted from the data, which were then
fit to the spectrum predicted by Eq.~(\ref{BR01}).  The data and the fit have been re-phased to
portray purely absorptive and dispersive quadratures.}}
\end{figure}

In driven-oscillation mode, a function generator was used to drive the EOM and the probe beam
optical rotation was detected with a lock-in amplifier (SR844, Stanford Research Systems, Inc.).
The modulation frequency was swept around 2$\omega_{\textrm{L}}$ in order to map out the $\Delta
m_F = 2$ magnetic-resonance curve, which was recorded on an oscilloscope. An example data set is
shown in Fig.~\ref{FO}. For these data, the pump beam power was set to 260 $\mu$W peak with a
50\% duty cycle and the probe beam was 55 $\mu$W continuous. Three resonances separated by the
$\sim$70 Hz nonlinear Zeeman splitting are clearly visible.

Due to electrical interference, a phase-coherent signal was picked up by the lock-in amplifier;
this produced a frequency-dependent offset even when the pump and probe beams were blocked. This
spurious baseline was subtracted from the data in Fig.~\ref{FO} and the data were fit to the
three-Lorentzian magnetic-resonance spectrum predicted by Eq.~(\ref{BR01}). According to this
fit, the central ($m_F = -1 \rightarrow m_F = +1$) magnetic resonance occurs at 682 504.318 $\pm$
0.050 Hz (1$\sigma$ uncertainty). Converting the best-fit frequency uncertainty into a field
uncertainty yields a magnetic sensitivity of 3.5 pT. The lock-in time constant was 10 ms and the
frequency sweep rate was 200 Hz/s, such that most of the spectrum was recorded within a span of
$\sim$2 seconds.  For the fitting procedure, the data were averaged in 10 ms bins in order to
reduce correlations in point-to-point noise which would erroneously reduce the estimated
frequency uncertainty.  As a cross-check of this sensitivity figure, many sets of simulated data
were generated with random noise which was statistically equivalent to the off-resonant noise
measured in the experiment. Repeated least-squares fitting of this simulated data yielded a
root-mean-square scatter of 0.046 Hz in best-fit frequency (equivalent to 3.3 pT) when all other
fitting parameters were held fixed.  The off-resonant noise indicates that if the EOM driving
frequency were set to the zero-crossing of the dispersive trace shown in Fig.~\ref{FO} and a
steady-state experiment performed, fluctuations of \lessgtrsim{\gtrsim}9.6 pT could be detected
with a 1 Hz noise bandwidth.


This magnetometric sensitivity was achieved in spite of several sub-optimal experimental
conditions: high pump and probe beam powers, excessive pump duty cycle, analog data transmission,
and sub-optimal field orientation\cite{Note3}. In a reference sensor consisting of identical
components interrogated non-remotely\cite{Corsini_PhD}, optimization of the pump power and duty
cycle yielded an optical rotation signal 14 times larger than that shown in Fig.~\ref{FO}.  This
implies that optimization of the pump beam characteristics could immediately yield sub-pT
sensitivity in the remote scheme. Moreover, these signals were recorded in an unshielded
environment and subject to fluctuations in the ambient field which were often larger than 10
pT/s. Most likely the sensitivity demonstrated here is limited by genuine field fluctuations.
Improved sensitivities can therefore be expected in future experiments, particularly if a
gradiometric scheme is employed.

In self-oscillating mode the polarimeter output was conditioned by a triggering circuit to drive
the EOM directly, generating a positive feedback loop and causing the system to oscillate
spontaneously at the magnetic-resonance frequency. A passive band-pass filter of width 10 kHz
centered around $\omega_{\textrm{L}}$ was included in the loop to reduce broadband noise fed into
the triggering circuit. (Oscillation at 2$\omega_{\textrm{L}}$ was also possible, but less robust
due to AM radio interference and the smaller signal amplitude.) The probe beam power was 50
$\mu$W leaving the launcher; the pump beam power was 10 $\mu$W time-averaged with a low (10-20\%)
duty cycle. To quantify the magnetometer's performance we mixed down its self-oscillation signal
with that of the reference sensor using the lock-in amplifier, with the reference sensor acting
as the external frequency reference and the remote sensor as the signal input.  Helmholtz coils
near the test sensor generated a field offset seen by the two magnetometers.  This gradient was
tuned to generate a self-oscillation beat frequency of $\sim$275 Hz and the lock-in time constant
set to 1 ms. The output of the lock-in amplifier was recorded with a data acquisition card and
saved to a computer.

\begin{figure}[t]
\includegraphics[scale=1]{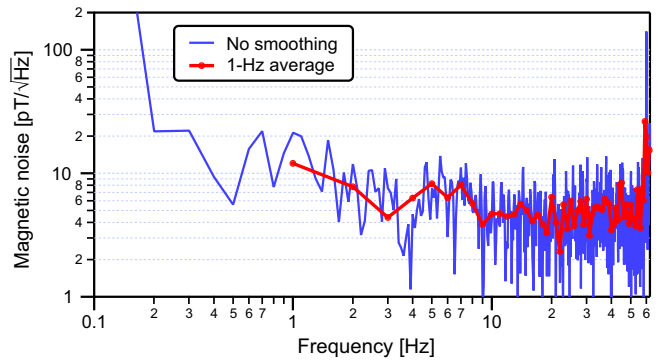}
{\caption[Magnetic-noise PSD.]{\label{PSD} (Color online) PSD of the gradiometer signal.  The
beat frequency of the sensors was calculated as a function of time, converted into a field
difference, and Fourier transformed to yield the magnetic noise as a function of frequency. The
thin blue trace is the Fourier transform; the thicker red trace is the same data smoothed into 1
Hz bins. The mean noise floor between 1 Hz and 50 Hz is 5.3 pT/$\sqrt{\textrm{Hz}}$. Ambient
60-Hz magnetic field noise can be clearly seen.}}
\end{figure}

Figure \ref{PSD} shows an analysis of the magnetic-field noise observed by the sensors.  The
SR844 output was digitally filtered with a 125 Hz band-pass filter about the intermediate
frequency, then fit with a running sine wave in 8 ms segments to calculate the beat frequency as
a function of time.  This frequency was then converted into a fluctuation about Earth's field
using Eq.~(\ref{BR01}), assuming that both magnetometers were oscillating on the same $\Delta m_F
= 1$ resonance\cite{Note4}. A Fourier transform of the field difference yielded the power
spectral density (PSD) of the reported magnetic noise. The average noise floor from 1 Hz to 50 Hz
was 5.3 pT/$\sqrt{\textrm{Hz}}$. It is not clear how much of the PSD noise floor arises from
sensor noise and how much can be attributed to current noise in the power supply driving the
Helmholtz coils or fluctuations in the ambient field gradient. A principal advantage of the
self-oscillating scheme is its high bandwidth -- AC magnetic fields of frequency
\lessgtrsim{\gtrsim}1 kHz and magnitude 1 nT have been detected with high SNR using a version of
the reference sensor described here.



Although a flat mirror was used in these experiments, such a configuration is not practical for
remote magnetometry because the mirror must be aligned at the cell in order to reflect the probe
beam back to the polarimeter.  Replacing this mirror with a polarization-preserving
retroreflector would allow for truly adjustment-free, long-baseline magnetometry. Corner-cube
reflectors can significantly alter the polarization properties of an interrogating laser. An
omnidirectional retroreflecting sphere is a promising choice, with a graded-index Luneberg-like
sphere being ideal. Polarimetry tests were conducted using a surrogate 1 cm diameter sphere with
index of refraction $n\approx2$, silver coated on its distal surface. Initial tests using a 633
nm laser and a digital polarimeter showed that the retroreflector (O'Hara Corp) preserved linear
polarization of a probe beam to within a few degrees of ellipticity, which represented the
measurement error of the polarimeter. Future tests will incorporate this retroreflector in the
magnetometer design.  In addition, the free-space baseline of the magnetometer will be increased,
with the expectation that this technique can be extended to distances of several hundred meters
before atmospheric seeing becomes a significant noise source\cite{Buck_AO67}. Beyond this
distance scale, adaptive optics techniques may become necessary to mitigate the effects of
atmospheric turbulence and retain magnetometric sensitivity.

A sensitive remote magnetometer capable of being interrogated over several kilometers of free
space would be highly desirable in several applications, including ordnance detection, perimeter
monitoring, and geophysical surveys.  Inexpensive manufacturing of the cell/retroreflector
package would allow many such sensors to be widely distributed and interrogated by a single
optical setup. Further research in remote magnetometry will also contribute to recently proposed
efforts to measure the Earth's magnetic field using mesospheric sodium atoms and laser guide-star
technology\cite{Higbie_PNAS11}.

The authors would like to thank collaborators Charles Stevens and Joseph Tringe (LLNL) for
initiating this project and providing the retroreflector,  as well as Mikhail Balabas for the
antirelaxation-coated $^{87}$Rb cells.  We also thank Mark Prouty, Ron Royal, and Lynn Edwards at
Geometrics, Inc.\ for experimental assistance and use of magnetometric facilities.  We
acknowledge support from Victoria Franques through the Department of Energy's National Nuclear
Security Agency (NNSA-NA-22)NA 22, Office of Nonproliferation Research and Development.  This
work was also supported in part by the Navy (contract N68335-06-C-0042), by the Department of
Energy Office of Nuclear Science (Award DE-FG02-08ER84989), and by NSF (ARRA 855552).  Parts of
this work performed under the auspices of the U.S. Department of Energy by Lawrence Livermore
National Laboratory under Contract DE-AC52-07NA27344.

\end{document}